\documentclass[prb,twocolumn,preprintnumbers,%
               showpacs,nofootinbib,floatfix]{revtex4}
               
\usepackage{amsmath}
\usepackage{amsfonts}
\usepackage{tikz}
\usepackage{xcolor}
\usepackage{amssymb}
\usepackage{amsthm}
\usepackage[english]{babel}
\usepackage{graphicx}
\usepackage{textcomp}
\usepackage{framed}
\usepackage{multirow}
\usepackage{bm}
\usepackage{float}
\usepackage[pdfauthor={G. Ceccarelli and J. Nespolo},
            pdftitle={Universal scaling of three-dimensional bosonic %
            gases in a trapping potential}]{hyperref}
\usepackage{microtype}

\newcommand{\avg}[1]{\langle #1 \rangle}
\newcommand{\helicity}{\ensuremath{{R_\Upsilon}}}
\newcommand{\ry}{\helicity}
\newcommand{\rxi}{\ensuremath{{R_\xi}}}
\newcommand{\tc}{\ensuremath{T_c}}

\renewcommand{\vec}[1]{\mathbf{#1}}

\begin{document}

\title{Universal scaling of three-dimensional bosonic gases in a trapping %
potential}

\author{{Giacomo Ceccarelli} and {Jacopo Nespolo}}
\affiliation{Dipartimento di Fisica dell'Universit\`a di
			 Pisa and INFN, Sezione di Pisa, I-56127 Pisa, Italy}

\date{\today}
\begin{abstract}
  We investigate the critical properties of cold bosonic gases in three
  dimensions, confined by an external quadratic potential coupled to the 
  particle density, and realistically described by the Bose-Hubbard (BH) model.
  The trapping potential is often included in experiments with cold atoms
  and modifies the critical finite-size scaling of the homogeneous system in a 
  non trivial way.
  The trap-size scaling (TSS) theory accounts for this effect through the
  exponent 
  $\theta$.
  
  We perform extensive simulations of the BH model at the critical 
  temperature, in the presence of harmonic traps.
  We find that the TSS predictions are universal once we account for the
  effective way in which the trap locally modifies the chemical potential 
  $\mu$ of the system.
  The trap exponent for the BH model at $\mu=0$ is the one 
  corresponding to an effective quartic potential.
  At positive $\mu$, evidence suggests that TSS breaks down sufficiently far 
  from the centre of the trap, as the system encounters an effective phase 
  boundary.
\end{abstract}


\maketitle

\section{Introduction}
A key aspect of the theory of critical phenomena is universality:
the relevant properties of physical systems at criticality depend only on some 
global features, such as the dimensionality and the invariance symmetries of 
the underlying Hamiltonian, while the microscopic details of the interaction 
play no role to this end.
The different physical systems can then be catalogued into universality
classes, according to their critical behaviour.
Consequently, models in the same class share the values of the critical 
exponents, as well as the shape of scaling functions.
Moreover, simplified theoretical models have predictive power on the universal
properties of complex experimental systems, as long as they fall in the same
universality class. \cite{sachdev2011}

In recent years, thanks to the great improvements in the experimental
capabilities of handling ultra-cold atoms, it has become possible to create
many-body systems which quite closely realise the theoretical models
commonly studied in condensed matter 
physics.\cite{Science.315.1556, RevModPhys.71.463, RevModPhys.80.885}
In particular, cooling techniques and optical lattices led to the experimental
observation of the Bose-Einstein condensation (BEC) and of the superfluid to
Mott-insulator\cite{Nature.415.39} quantum phase transition in lattice 
bosonic gases experiments.

The Bose-Hubbard (BH) model\cite{PhysRevB.40.546} provides a realistic
description for these experimental systems\cite{PhysRevLett.81.3108}. 
It is defined by the Hamiltonian
\begin{align}
 H_{\rm BH} = &- \frac{J}{2} \sum_{\langle \vec{xy}\rangle}
                 \left (b_{\vec{x}}^{\dag} b_{\vec{y}}^{} +
                 b_{\vec{y}}^{\dag}b_{\vec{x}}^{}\right)
                 - \mu \sum_{\vec{x}} n_{\vec{x}} \nonumber \\
              &+ \frac{U}{2} \sum_{\vec{x}} n_{\vec{x}} (n_{\vec{x}} - 1) \; ,
\end{align}
where $n_{\vec{x}} = b_{\vec{x}}^{\dag} b_{\vec{x}}^{}$ is the particle density
operator, $b_{\vec{x}}^{\dag}$ is the bosonic creation operator and
$\langle \vec{xy}\rangle$ indicates nearest neighbour sites on a cubic lattice.
The chemical potential $\mu$ acts as a control parameter coupled to the 
particle density and $U > 0$ is the strength of the contact repulsion between
particles.
In the following, we take the hard-core (HC) limit $U \to \infty$.
In this way the local particle number operator is restricted to the values
$\lbrace 0, \, 1\rbrace$ only.
In the spirit of universality, this change only affects the strength of the
interaction among atoms, and it is not expected to change the universal
behaviour of the model near phase transitions.

The theory of phase transitions generally applies to homogeneous systems in the
thermodynamic limit, so that the comparison with experiments may not be 
immediate.
In particular current experiments almost always include a trapping
potential\cite{Grimm200095} to keep the atoms confined on a limited region
of the optical lattice.
The shape of the trap is usually well approximated by a power-law profile of
the form
\begin{equation}\label{eq:potential}
V(\vec{r}) = v^p r^p ,
\end{equation}
where the parameter $v$ is related to the strength of the confinement and $p$
is a positive integer. Generally, parabolic traps are used ($p=2$).
The presence of the trap modifies the critical behaviour of the gas: most
notably, the correlation length $\xi$, which usually diverges at phase
transitions, is bound to remain finite by the confining potential, so that the
(homogeneous) phase transition gets suppressed in the (inhomogeneous) real
system.
Only in the limit $v \to 0$, in which the trap is switched off, a true phase
transition is expected to be seen.

It is then evident that a correct modelling of the experimental setup must
include the effect of the confinement and, to this end, the trapping
interaction must be added to the fundamental Hamiltonian of the systems under
investigation.
Since this interaction rules out the phase transition, in the language of the
renormalisation group (RG) theory, it constitutes a new relevant field, to 
which a new critical exponent is expected to be associated.
The theoretical setting needed to investigate the phenomenology of phase
transitions in the presence of a trap is the trap-size scaling (TSS) theory.
\cite{PhysRevLett.102.240601, PhysRevA.81.023606}
In this framework, the effect of the trap on the critical behaviour is encoded
in the trap critical exponent $\theta$.
The physical meaning of $\theta$ can be understood noting that, when the
external parameters of the system are tuned to the values of the homogeneous
phase transition, the correlation length of the trapped system scales as
$\xi \sim l^\theta$, where we defined the trap size
\begin{equation}
l \equiv \frac{J^{1/p}}{v}
\end{equation}
which is the natural length scale associated to the potential.
In the following we fix the energy units setting $J=1$.
Of course $\xi$ diverges in the $l \to \infty$ limit, as we expect at a true
phase transition.

\begin{figure}
\includegraphics[width=8.5cm]{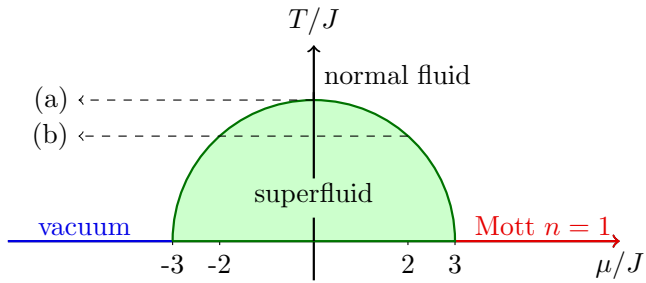}
\caption{(Colour online) A qualitative sketch of the phase diagram of the 
         three-dimensional hard-core Bose-Hubbard model. 
         The dashed lines represent the local effective potential 
         $\mu_{\rm eff}(r)$ starting from the critical point at $\mu=0$ (a) 
         and at $\mu=2$ (b).
         The distance from the centre of the trap increases in the direction
         of the arrow.
         Notice how the line (b) crosses the phase boundary at $\mu=-2$.}
\label{fig:phase-space}
\end{figure}

The physical problem we want to analyse in this paper concerns the issue of
the universality of the modified critical behaviour.
Following the RG ideas, we expect that the critical properties of trapped
systems, summarised in the exponent $\theta$, depend only on some global
and general features, such as the way the potential is coupled to the
critical modes of the unconfined system, the shape of the potential and the
homogeneous universality class.
To this end, we analyse the three-dimensional (3D) BH model, which belongs to
the homogeneous 3D XY universality class, in the HC limit and at
the finite-temperature phase transition from a normal fluid to a superfluid 
(see phase diagram in Fig.~\ref{fig:phase-space}).
The order parameter of this transition is the phase of the condensate wave
function, which is related to superfluid density.
The critical exponents for this transition are $\nu = 0.6717(1)$ and
$\eta = 0.0381(2)$.\cite{PhysRevB.74.144506}

We recall that phase transitions at $T=0$ are driven by quantum fluctuations, 
whose ultimate origin is the Heisenberg uncertainty principle.
On the other hand, finite-$T$ transitions are always classical in nature: 
the occupation number of the states corresponding to the critical modes
diverges, so that classical statistics may apply.
The universality class of the $U(1)$-invariant \emph{quantum} BH model at 
finite-$T$ is then the one of the \emph{classical} XY model.

When the confining potential is turned on, we consider the full system 
Hamiltonian 
\begin{equation}\label{eq:bh_trap}
 H = H_{\rm BH} + \sum_{\vec{x}} V(\vec{x}) n_{\vec{x}},
\end{equation}
where the trapping potential is coupled to the particle density.
A standard RG analysis within the TSS framework leads to the 
result\cite{PhysRevB.87.024513}
\begin{equation}
\label{eq:theta}
 \theta(p) = \frac{p\nu}{1+p\nu}
\end{equation}
for the relevant critical exponent $\theta$.
The TSS of the 3D HC BH model has been investigated in 
Ref.~\onlinecite{PhysRevB.87.024513} for the range $-3<\mu<0$.
At $\mu=-2$ it was found $T_c=0.7410(1)$ and the scaling functions of the
microscopic degrees of freedom are ruled by the exponent $\theta=0.57327(4)$,
which was obtained from the basic RG prediction (\ref{eq:theta}) using the known
value of $\nu$ from Ref.~\onlinecite{PhysRevB.74.144506}.
In a recent work on the 2D HC BH model \cite{PhysRevB.88.024517},
we proposed that the universal critical features of phase transitions in a
trapped system do not only depend on the bare shape of the confining potential
but also on the particular way in which the trap locally modifies the control
parameter $\mu$.
The actual local phase space position of the system can then be tracked by
means of an effective chemical potential $\mu_{\rm eff}(\vec{r})$.
In the following we show that the TSS scaling of the 3D HC BH model for a
chemical potential $\mu \neq 0$ follows the standard $\theta(p)$
behaviour \ref{eq:theta}, whereas the correct scaling at $\mu = 0$ is given by
$\theta(2p)$.
Furthermore, the scaling functions are different depending on the sign of the
chemical potential.
In particular, in the $\mu>0$ conditions, the system falls in an effective
superfluid phase up to a distance $r_{\rm bd}$ from the centre of the trap, at
which TSS breaks down.

The paper is organised as follows.
For the study of criticality in the presence of a trap, a very precise
determination of the homogeneous parameters at the phase transition point is
needed.
This is because we want to analyse the emergence of the known homogeneous
behaviour in the limit $l \to \infty$ which removes the trap.
The measurement of the transition temperature at $\mu = 0$ is reported in
detail in Sec.~\ref{sec:homogeneous}.
In Sec.~\ref{sec:trapped} we examine the model in the presence of the external
trapping potential:
we verify our findings by performing a trap-size scaling (TSS) analysis of the
correlation function and a finite-size trap-size scaling (FTSS) study of 
suitable observables, both at zero chemical potential and for the $\mu > 0$ 
case.
Finally, in Sec.~\ref{sec:conclusion} we discuss the main results of the
present work and draw our conclusions.

\section{Homogeneous system}
\label{sec:homogeneous}
In order to perform a detailed FSS analysis of the homogeneous model and 
determine its critical temperature, we consider the helicity modulus $\Upsilon$
and the second moment correlation length $\xi$.

The helicity modulus $\Upsilon$ is defined as
\begin{equation}\label{eq:helicity_def}
 \Upsilon \equiv - \frac{1}{L} \left.
 \frac{\partial^2 Z(\phi)}{\partial\phi^2}
 \right|_{\phi = 0} \; ,
\end{equation}
where $Z$ is the partition function under a twist $\phi$ of the boundary
conditions in one direction.\cite{PhysRevA.8.1111}
In our QMC simulations the quantity $\Upsilon$ is simply 
related\cite{PhysRevB.56.11678} to the
linear winding number $W$ through the relation
\begin{equation}
 \Upsilon = \frac{\langle W^2 \rangle}{L} \; .
\end{equation}

The two-points Green function $G_b(\vec{x},\vec{y})$ is defined as
\begin{equation}\label{eq:G_homo}
 G_b(\vec{x},\vec{y}) = \langle b^{\dag}_{\vec{x}} \, b_{\vec{y}} \rangle.
\end{equation}
The homogeneous system with periodic boundary conditions is translational 
invariant, so that the Green function only depends on the separation 
$\vec{r}= \vec{x} - \vec{y}$ between the two points. 
We can thus restrict the study to
$G_b(\vec{r}) \equiv G_b(\vec{r},\vec{0})$.
Finally, we denote the lattice Fourier transform of $G_b(\vec{r})$ as
${\widetilde{G}_b(\vec{p})}$.
The second moment correlation length $\xi$ is then defined 
as\cite{PhysRevB.87.024513}
\begin{equation}
\label{lunghezza}
 \xi^2 \equiv  \frac{1}{4 \sin^2 (\pi/L)} 
 \frac{\widetilde{G}_b({\bf 0}) - \widetilde{G}_b({\bf p})}
      {\widetilde{G}_b({\bf p})},
\end{equation}
where $\vec{p}=(2\pi/L,0,0)$.

The quantities $\helicity = \Upsilon L$ and $\rxi = \xi/L$ are dimensionless
and RG invariants. 
For small $\tau \equiv T/\tc - 1 $, they follow the universal scaling relation
\cite{PhysRevB.87.024513}
\begin{equation}\label{eq:Rscaling}
 R = f(\tau L^{1/\nu}) + L^{-\omega} f_\omega(\tau L^{1/\nu}).
\end{equation}

\subsection{FSS analysis for the homogeneous system}
\label{sec:FSS}
We performed quantum Monte Carlo (QMC) simulations of the 3D HC BH model, for
lattice sizes up to $L=32$.
Within the stochastic series expansion framework\cite{PhysRevB.43.5950}, 
we use the directed operator-loop algorithm 
\cite{PhysRevE.66.046701, PhysRevE.64.066701}.
More details on our implementation of the QMC can be found in 
Refs.~\onlinecite{PhysRevA.85.053637, PhysRevA.85.023616}.
Our simulations for the homogeneous system are approximately $4 \times 10^6$ 
Monte Carlo steps (MCS) long. 
We decorrelated the data by applying the blocking method and the errors are 
calculated through a jackknife analysis.
For a discussion of the self-correlation times of the data, see
Appendix~\ref{sec:selfcorr}.

\begin{figure}[tb]
 \centering
 \includegraphics[width=8.5cm,keepaspectratio=true]{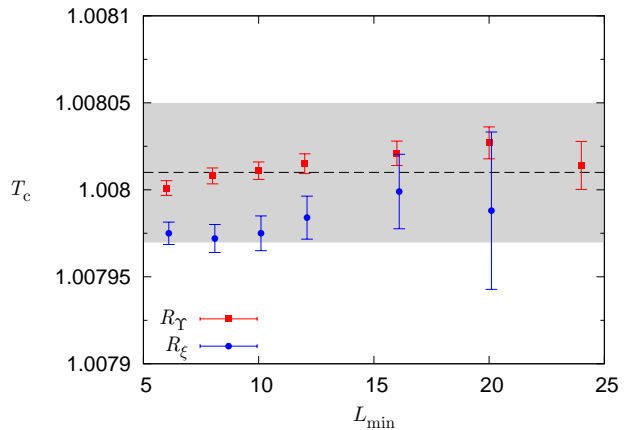}
 \caption{(Colour online) Estimates of $T_c$ obtained from the FSS analysis of
          \ry\ and \rxi, considering data with $L\geq L_{\rm min}$.
 }
 \label{fig:tc}
\end{figure}

Close to the asymptotic regime, Eq.~\ref{eq:Rscaling} can be expanded as a
Taylor series about $\tau = 0$:
\begin{equation}\label{eq:R_expansion}
 R = R^* 
     + \sum_{i=1}^n a_i \tau^i L^{i/\nu} 
     + L^{-\omega} \sum_{j=0}^m b_j \tau^j L^{j/\nu}.
\end{equation}
The asymptotic values for the helicity modulus and the correlation length are
known from previous works \cite{PhysRevB.74.144506},
\begin{equation}\label{eq:R_asy}
 \ry^* = 0.516(1), \qquad \rxi^* = 0.5924(4),
\end{equation}
as are the exponents $\nu$ and $\omega$,
\begin{equation}\label{eq:nu_omega}
 \nu = 0.6717(1), \qquad \omega=0.785(20).
\end{equation}
The data for \ry\ and \rxi\ at different $L$ and $T$ can then be fitted against
the first few terms of this expansions.
The optimal number of terms to use in the fit (i.e.~$m$ and $n$ in 
Eq.~\ref{eq:R_expansion}) is determined by progressively adding more terms to 
the series and looking for the stabilisation of the fit parameters and to when 
the residuals start degrade.
Residual corrections to scaling are assessed by repeating the fit discarding
the data for lattice sizes $L < L_{\rm min}$ while progressively increasing
$L_{\rm min}$.

For the fit of \ry\ data, we found optimal to use $n=1$ and $m=0$,
while the analysis of \rxi\ requires higher order corrections
$O(L^{-2\omega})$.
The data used in the analyses was chosen in a self-consistent way, by
only retaining the data points satisfying
\begin{equation}\label{eq:data_selection}
 \left| R / R^* - 1 \right| \leq 0.1.
\end{equation}
The limit of 10\% deviation from the asymptotic value in the formula above was
set by requiring that the $\chi^2$ be acceptable.
The results of the fits on \ry\ and \rxi\ are reported in
Table~\ref{tab:tc_fit} and plotted in Fig.~\ref{fig:tc}. 
The scaling ansatz of Eq.~\ref{eq:R_expansion} accurately accounts for 
subleading corrections to scaling.
This, together with the self-consistent choice of the fitting window, 
Eq.~\ref{eq:data_selection}, allows us to obtain a precise estimate of the 
critical temperature with simulation data on relatively small lattice sizes.

The analyses on both observables converge to a common value.
Our final estimate for the critical temperature of the 3D HC BH model at
$\mu=0$ is
\begin{equation}\label{eq:tc_mu0}
 \tc^{(\mu = 0)} = 1.00801(4).
\end{equation}
We consider the value of \tc\ extracted from the fit on \ry\ to be more
reliable, due to the stability of the observable and to the residual $\chi^2$
obtained.
The results from \rxi\ are a cross-check and we use them to better estimate
the error on \tc.
The latter must also take into account the uncertainties on the other 
parameters entering in Eq.~\ref{eq:R_expansion}, namely those reported in
Eqs.~\ref{eq:R_asy} and \ref{eq:nu_omega}.
A standard bootstrap analysis shows that the error introduced by these
quantities is not negligible, yet it decreases for increasing lattice size.
For the fits on \ry(resp.~\rxi) data and lattice sizes
$L\geq 10$ this error ranges between $1.5\div0.8 \times 10^{-5}$ 
(resp.~$0.5\div0.3 \times 10^{-5}$). 
The quoted error $\Delta \tc = 4 \times 10^{-5}$ 
accounts for all of these effects. 

Our value of \tc\ agrees with the previous estimates of 
Ref.~\onlinecite{EuroPhysLett.99.66001} and 
Ref.~\onlinecite{PhysRevA.86.043629}, quoting respectively
$\tc = 1.008(3)$ and $\tc = 1.00835(25)$.

\begin{table}
\caption{Estimates of \tc\ with corresponding statistical error.}
\label{tab:tc_fit}
\begin{ruledtabular}
\begin{tabular}{rlllll}
$L_{\rm min}$ & $\tc^{(\Upsilon)}$ %
        & $\chi^{2(\Upsilon)}_{/{\rm dof}}[{\rm dof}]$ & $\tc^{(\xi)}$ %
& $\chi^{2(\xi)}_{/{\rm dof}}[{\rm dof}]$  \\
\hline
5    & 1.007984(4)   & 8.5[31]  & 1.007980(6)   & 2.9[39] \\
6    & 1.008001(4)   & 2.4[27]  & 1.007975(6)   & 2.1[34] \\
8    & 1.008008(5)   & 1.8[23]  & 1.007972(8)   & 2.1[29] \\
10   & 1.008011(5)   & 1.8[19]  & 1.00798(1)    & 2.2[24] \\
12   & 1.008015(6)   & 1.8[15]  & 1.00798(1)    & 2.2[19] \\
16   & 1.008021(7)   & 1.7[11]  & 1.00800(2)    & 2.8[14] \\
20   & 1.008027(9)   & 1.9[7]   & 1.00799(5)    & 1.6[7]  \\
24   & 1.00801(1)    & 1.2[3]   & --            & --      \\ 
\end{tabular}
\end{ruledtabular}
\end{table}

\section{Trapped system}
\label{sec:trapped}

The question we want to address in this paper is related to the universality of
the TSS theory.
The trap exponent $\theta$ depends on the power $p$ of the trapping potential,
cf.\ Eq.~\ref{eq:potential}.
We suggested in Ref.~\onlinecite{PhysRevB.88.024517} that the trap exponent
$\theta$ predicted by TSS is indeed universal throughout the 3D XY universality
class.
However, the particular shape of the BH phase diagram leads to a modified TSS
behaviour when $\mu = 0$. 
In this condition, the trap exponent is the one corresponding to a trapping
potential of power $2p$.

We recall that the trap-size limit is defined as the limit in which
$r,l \rightarrow \infty$ while keeping the ratio $\zeta=r/l^{\theta}$ fixed.
In this limit the argument of the trapping potential $r/l=\zeta/l^{1-\theta}$
vanishes, since $\theta<1$, so that only the short range behaviour is relevant
for the scaling features of the model.

The trapping potential couples to the density operator, and can thus be thought
as a {\it local} effective chemical potential
\begin{equation}
\mu_{\rm eff}(r) \equiv \mu - V(r).
\end{equation}
Calling $T_c(\mu)$ the critical temperature of the homogeneous system, 
we can define an effective temperature 
$T_{\rm eff}(r) \equiv T_c[\mu_{\rm eff}(r)]$. 
This is the temperature at the phase transition of a homogeneous 
system whose chemical potential is set to the value of $\mu_{\rm eff}$ at site
$r$ of the inhomogeneous system.
We argue that the critical modes of the inhomogeneous system can be
described by means of the local control parameter
\begin{equation}
\label{eq:tau_eff}
\tau_{\rm eff}(\mu,r) \equiv T_c(\mu) - T_{\rm eff}(r).
\end{equation}
One should keep in mind that the system is considered at equilibrium at 
the critical temperature. 
Here $\tau_{\rm eff}$ should be considered as an effective distance from the 
phase boundary, much in the same way as is $\tau$ [defined before 
Eq.~\ref{eq:Rscaling}]. 
However, $\tau$ bears a precise physical meaning, whereas $\tau_{\rm eff}$ is 
only a practical tool to describe the trapped critical behaviour.

Recalling that in the trap-size limit only the short-$r$ behaviour of the
trapping potential is relevant, we can expand the function $T_{\rm eff}$ in
Eq.~\ref{eq:tau_eff} around $\mu$, obtaining the general expression
\begin{equation}\label{eq:tau_mu}
\tau_{\rm eff}(\mu, r) \simeq T'_c(\mu) V(r) - \frac{1}{2} T''_c(\mu) V(r)^2.
\end{equation}

For $\mu \neq 0$, the first term, of order $r^p$, dominates the expansion.
We then expect the TSS behaviour of the 3D BH to agree with that of the trapped
3D XY universality class with the same trapping exponent $\theta(p)$.
However, for $\mu = 0$, the phase diagram of Fig.~\ref{fig:phase-space} tells 
us that the first derivative vanishes, and the second term, of order 
$r^{2p}$, becomes dominant. For this reason we expect that, at $\mu = 0$,
the critical behaviour be ruled by the exponent
\begin{equation}
\theta(2p) = \frac{2p\nu}{1+2p\nu},
\end{equation}
i.e., the system behaves as a classical 3D XY model trapped by a potential 
$U\sim r^q$ with exponent $q = 2p$.

In the presence of a trapping potential, the translational invariance is
broken.
Due to the spherical symmetry of the potential, it is then natural to replace
the two-point function \ref{eq:G_homo} with the correlation function with
respect to the centre of the trap,
\begin{align}
G_b(\vec{0}, \vec{r}) &\equiv  \langle b_\vec{0}^\dag b^{}_\vec{r} \rangle 
                 \nonumber \\
                &\approx l^{-(1+\eta)\theta}
                         \mathcal{G}_b(rl^{-\theta}, \tau l^{\theta/\nu}),
                         \label{eq:gr_tss}
\end{align}
where $\mathcal{G}_b$ is a universal function.
The inhomogeneous susceptibility is defined as 
\begin{equation}
 \chi_t \equiv \sum_\vec{x} G_b(\vec{0},\vec{x})
 \label{chi}
\end{equation}
and the second moment correlation length as
\begin{equation}
 \xi_t^2 \equiv \frac{1}{6 \chi_t} \sum_\vec{x} |\vec{x}|^2 
      G_b(\vec{0},\vec{x}) \; .
 \label{xi-trap}
\end{equation}
Note that $\chi_t$ is related only to the integral of the correlation
with the centre of the trap, and thus differs from the usual definition of the
susceptibility for the homogeneous system.

\begin{figure}[t!]
 \centering
 \includegraphics[width=8.5cm,keepaspectratio=true]{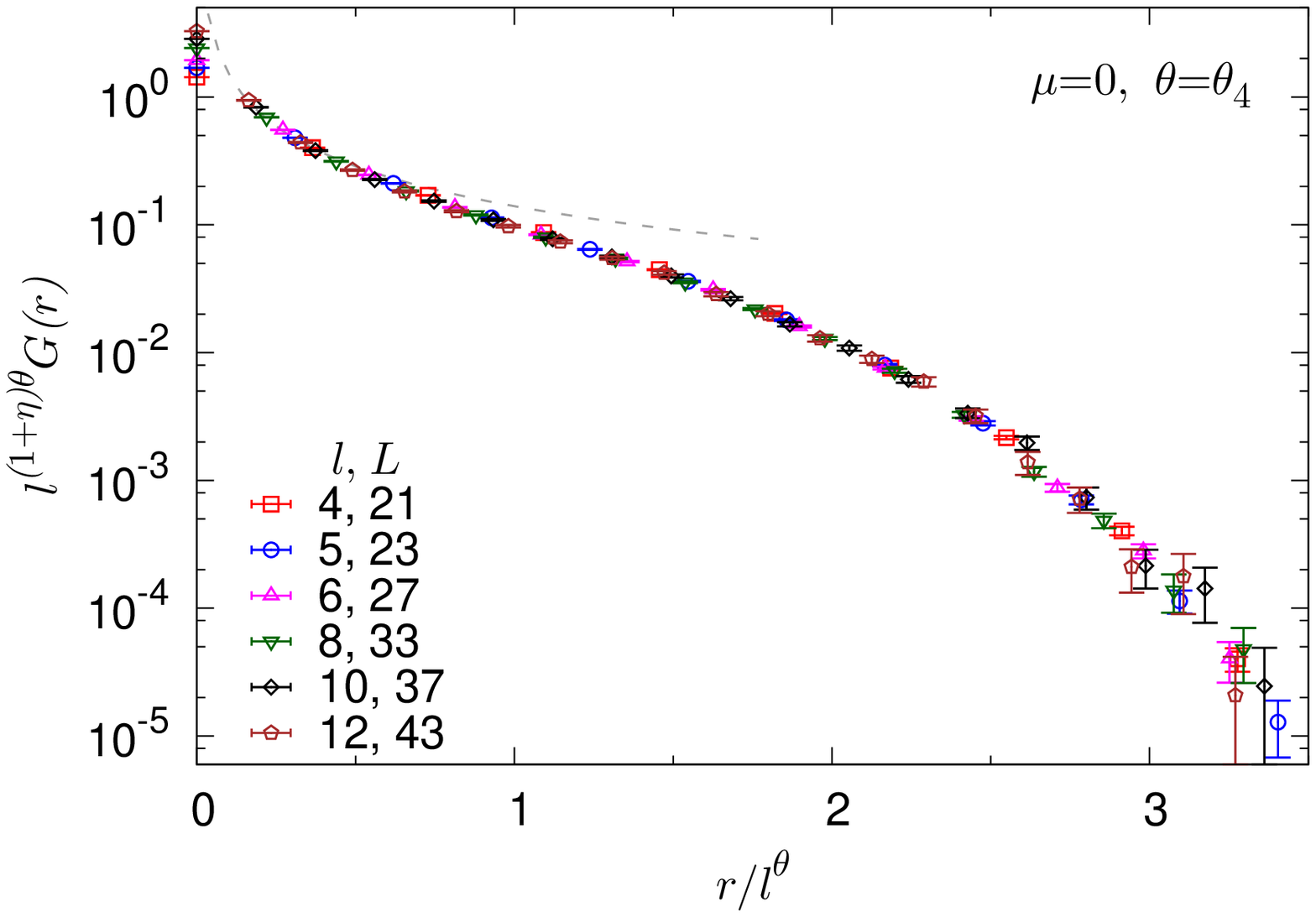}
 \includegraphics[width=8.5cm,keepaspectratio=true]{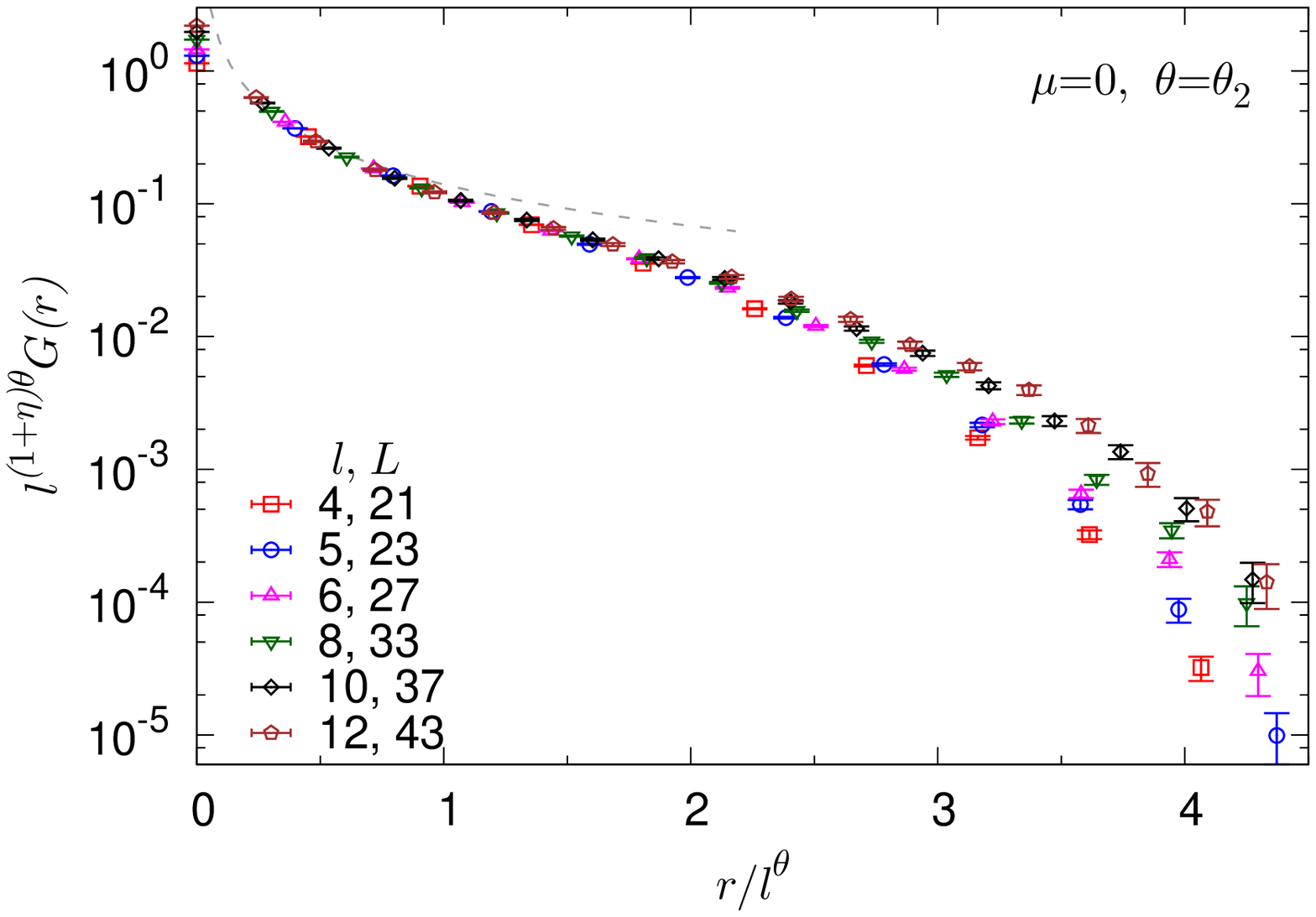}
 \caption{(Colour online) Scaling at $\mu = 0$ and $T=T_c^{\mu=0}$ of the 
          two-point function between the centre of the trap and points at
          distance $r$. 
          The exponents $\theta_4$ (top) and $\theta_2$ (bottom) are used.
          The homogeneous scaling (dashed line) holds close to the centre of the
          trap.}
 \label{fig:gr_mu0}
\end{figure}

\begin{figure}[t!]
 \includegraphics[width=8.5cm,keepaspectratio=true]{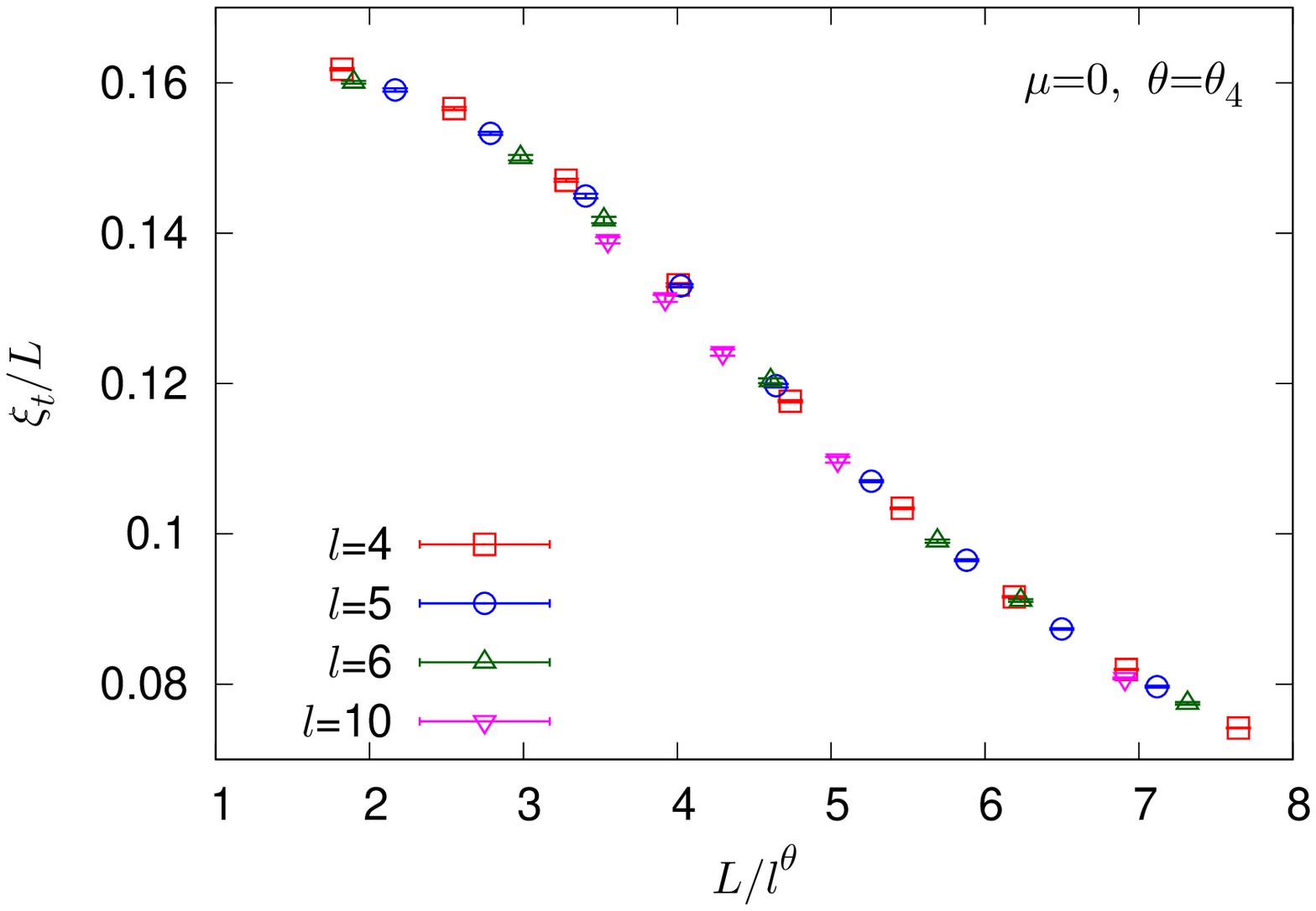}
 \includegraphics[width=8.5cm,keepaspectratio=true]{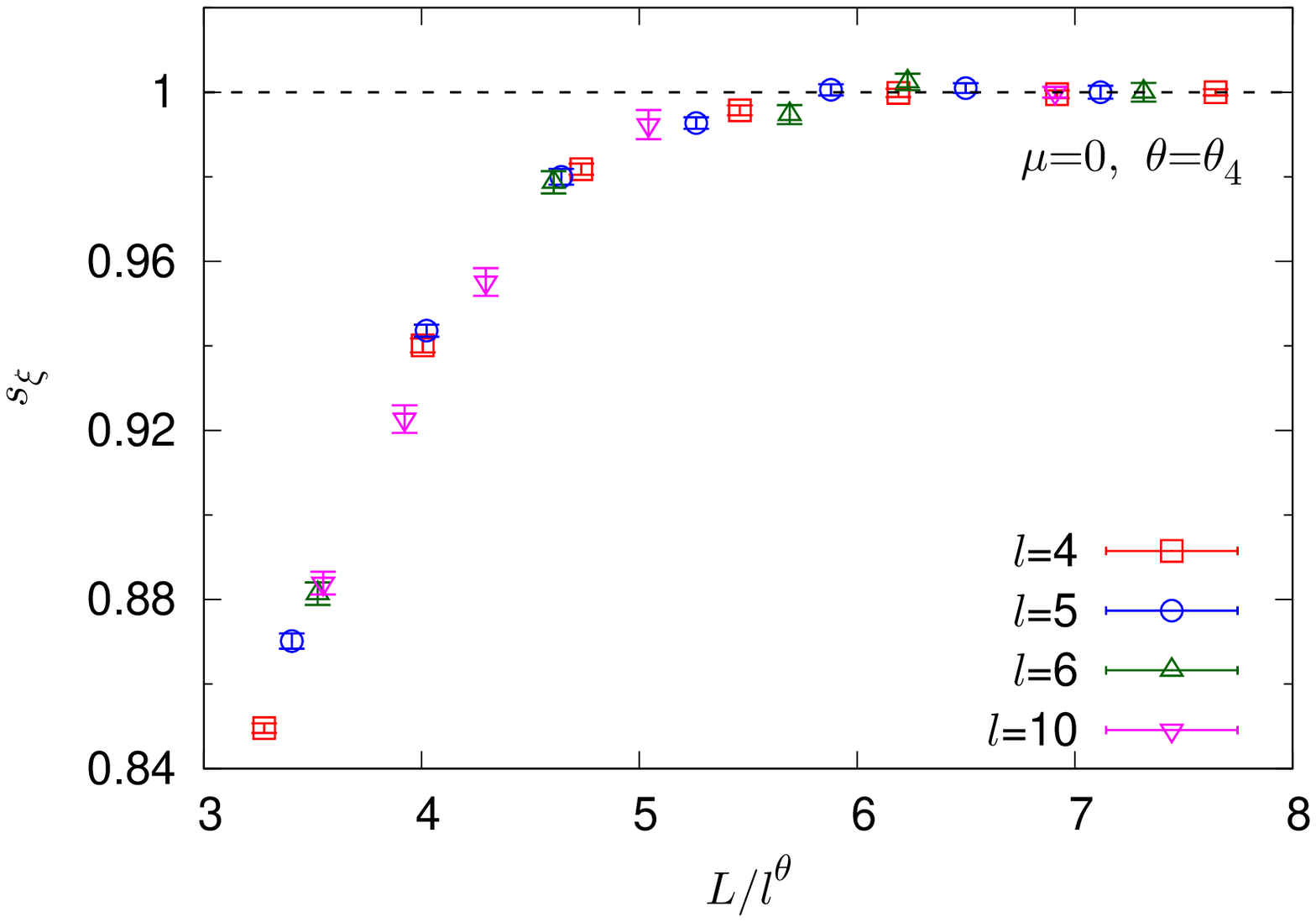}
 \caption{(Colour online) Finite-size trap-size scaling at $\mu=0$ and 
          $T = T_c^{\mu=0}$ for $\xi_t/L$ (top) and saturation curve of 
          $\xi_t$ (bottom).}
 \label{fig:ftss_xi}
\end{figure}

In our simulations of trapped systems, the trap is enclosed within a hard 
walled cubic box.
The size of the box $L$ is always an odd integer, so that the centre of the
trap falls exactly on top of the central site of the cubic box.
The size of the trap $l$ and that of the box $L$ both affect the critical
properties of the system, requiring a simultaneous finite-size and
trap-size analysis.
The following behaviours for the correlation length and the susceptibility are
expected:\cite{PhysRevE.81.051122}
\begin{equation}
 \label{eq:xi_chi_ftss}
 \xi_t = L \mathcal{R}(\tau l^{\theta/\nu}, L/l^\theta), \quad 
 \chi_t = L^{2-\eta} \mathcal{X}(\tau l^{\theta/\nu}, L/l^\theta).
\end{equation}
As discussed above, the exponent $\theta$ in these equations is 
\begin{align}
 \theta_2 &\equiv \theta(2) = 0.57327(4) \qquad \mbox{at }\mu\neq0,\\
 \theta_4 &\equiv \theta(4) = 0.72876(3) \qquad \mbox{at }\mu=0.
\end{align}

\subsection{TSS at $\mu=0$}
\label{sec:TSS_mu0}
\begin{figure}[t]
 \includegraphics[width=8.5cm,keepaspectratio=true]{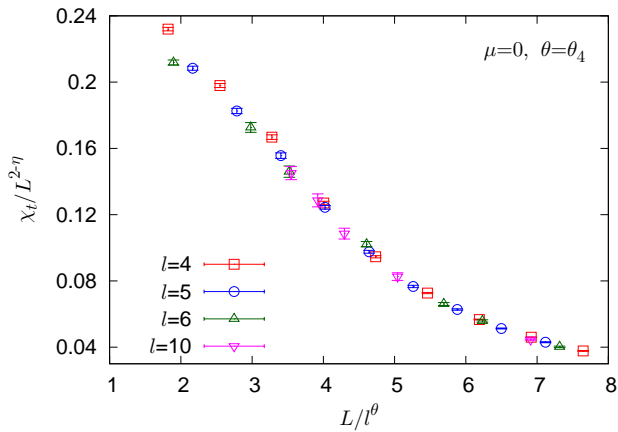}
 \caption{(Colour online) Finite-size trap-size scaling of $\chi_t$ at $\mu=0$
          and $T = T_c^{\mu=0}$.}
 \label{fig:ftss_chi}
\end{figure}

We simulated the model at $\mu=0$ and at the homogeneous critical temperature
of Eq.~\ref{eq:tc_mu0} for different trap sizes $l$ and lattice sizes $L$.
In the asymptotic condition $L \gg l^{\theta}$ it is possible to perform a pure
TSS study of the two-point correlation function $G_b(r)$.
The latter is a standard physical observable where it is possible to
check the validity of our reasoning.
In Figure~\ref{fig:gr_mu0}, we plot the data for $G_b(r)$ using the TSS ansatz
\ref{eq:gr_tss}, in which we set $\tau = 0$.
The data were generated keeping $L/l^\theta \approx 7$ (justified below) and
runs of approximately $(0.6 \div 1) \times 10^6$ MCS were used.
Here and below, the analysis method is analogous to the one used in the
homogeneous case discussed in Sec.~\ref{sec:homogeneous}, as are the
considerations related to self-correlation times.
Notice that, for $r \to 0$, the trap is locally flat, hence we expect to 
recover the homogeneous scaling, while only for $r>l^{\theta}$ does the effect 
of the trap becomes evident.
In this region, the rescaling of the correlation function plotted in 
Fig.~\ref{fig:gr_mu0} nicely supports the scaling with exponent $\theta_4$ (top
panel) against $\theta_2$ (bottom).

To further check our scaling predictions,
in Figures~\ref{fig:ftss_xi}-\ref{fig:ftss_chi}
we show the FTSS analyses.
The data presented in these figures come from QMC runs approximately 
$2 \times 10^5$ MCS long. 
Figure \ref{fig:ftss_xi}-(top panel) shows the rescaling of $\xi_t/L$: all the
data fall onto a single universal curve when using the predicted exponent
$\theta_4$.
The data for the observable $\chi_t/L^{2-\eta}$ confirm our claims and are  
shown Figure~\ref{fig:ftss_chi}.
Both Fig.~\ref{fig:ftss_xi} and Fig.~\ref{fig:ftss_chi} show small corrections
to scaling for low values of $L/l^\theta$.
The source of these discrepancies is to be found in the non-analytic
corrections due to irrelevant perturbations.
In Figure~\ref{fig:ftss_xi}-(bottom) we plot $\xi_t$ normalised with its 
asymptotic value: $s_\xi \equiv \xi_t(L)/ \xi_t(\infty)$. 
Operatively, we simulated the system at fixed trap size $l$ and increased the
lattice size $L$ until saturation; at this point, the value of the observable
at the largest $L$ was used as $\xi_t(\infty)$ to fix the normalisation.
From this figure we observe that the data saturate for $L/l^\theta \gtrsim 6$,
indicating that, for larger lattice sizes $L$ (as was the case for the
previous analysis of $G_b$), the hard-walled box does not influence the
scaling inside the trap.
The data for $\chi_t$ agree with these considerations.
Both the data for $\xi_t/L$ and $\chi_t/L^{2-\eta}$, when rescaled with 
the wrong exponent $\theta_2$, do not collapse onto a single curve, similarly 
to what is shown in Fig.~\ref{fig:gr_mu0}-(bottom).
We conclude that our QMC results are consistent with the scaling prediction of
the preceding section and discriminate between the two exponents, 
$\theta_2$ and $\theta_4$.

To conclude this section, we point out that similar scaling relations hold for
the density-density correlator.
However, this correlator is significantly different from zero only in a very
narrow region around the centre of the trap. 
To have acceptable signal to noise ratios, larger values of $l$ are needed,
whose computational cost makes them impractical to simulate.

\subsection{FTSS at $\mu>0$}
\label{sec:FTSS_mu2}
\begin{figure}[b!]
 \centering
 \includegraphics[width=8.5cm]{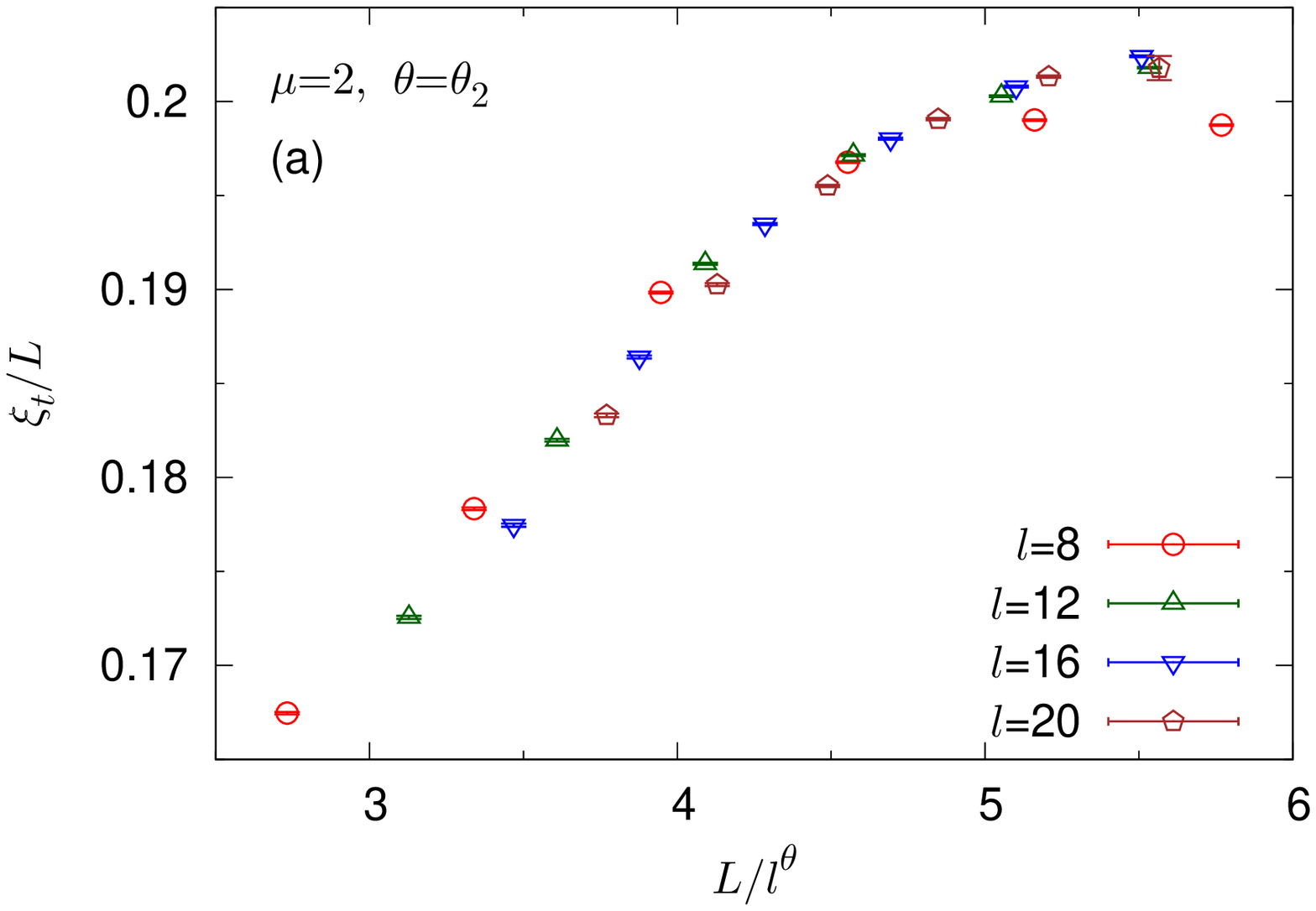}
 \includegraphics[width=8.5cm]{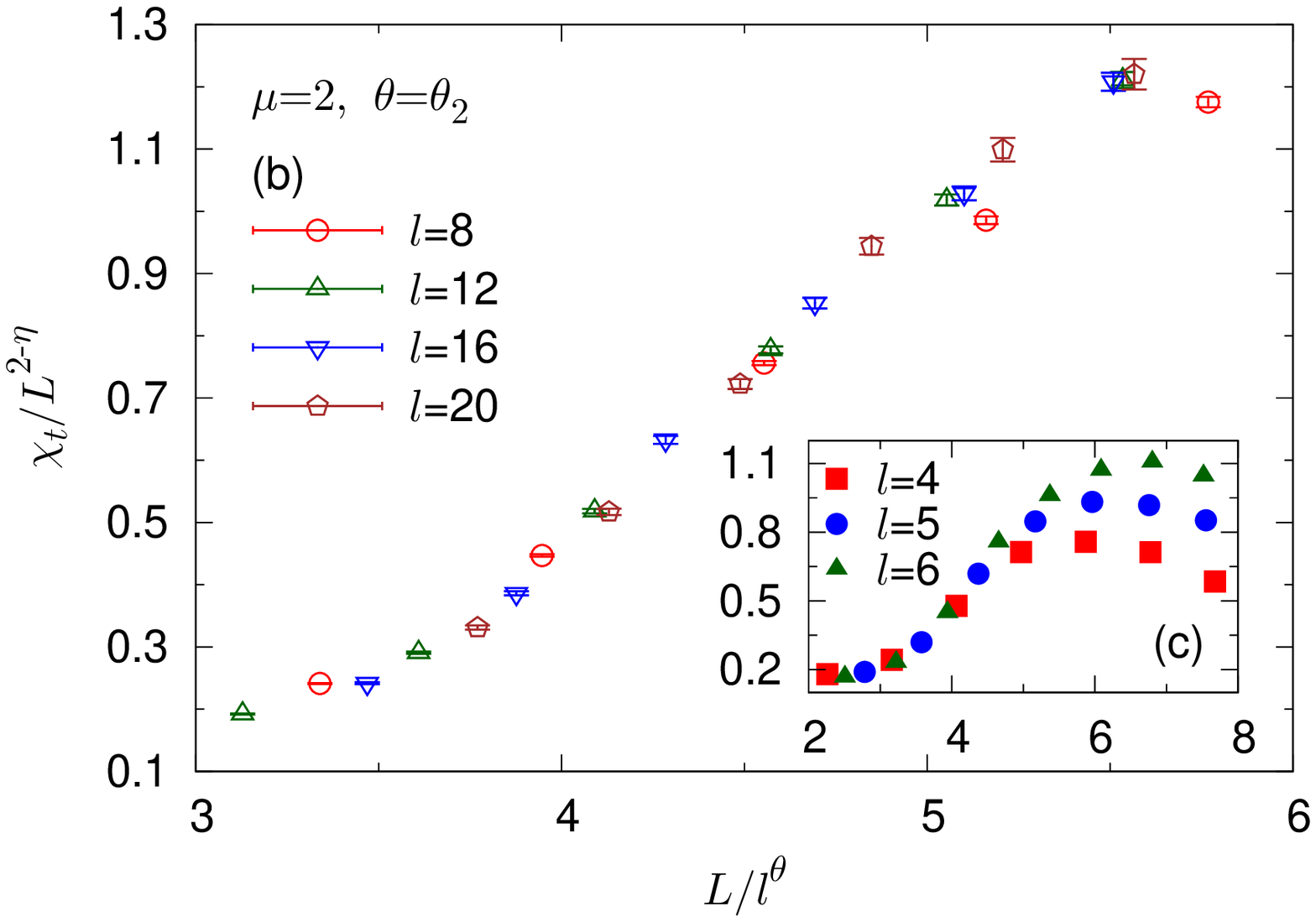}
 \caption{(Colour online) FTSS plots of $\xi_t/L$ (a) and $\chi_t/L^{2-\eta}$
          (b) at $\mu=2$ and $T=T_c^{\mu=2}$ using the predicted exponent 
          $\theta_2$.
          The curves, especially those for $\xi_t/L$ (a), are affected
          by large corrections to scaling.
          For large $L/l^\theta$, each fixed-$l$ curve tends to abandon
          the asymptotic curve, indicating the break down of the FTSS close to
          the effective phase boundary crossing (c).
          }
 \label{fig:xi_chi_mu-2}
\end{figure}

Having verified that the exponent at $\mu=0$ is the one expected for the
effective quartic potential, we now need to check that at $\mu \neq 0$ the
scaling behaviour is determined by the exponent $\theta_2$ corresponding to the
harmonic trap.
A previous work\cite{PhysRevB.88.024517} investigated the model at $\mu = -2$
and already confirms the theory.
In that case, moving away from the centre of the trap, i.e. decreasing the
effective chemical potential $\mu_{\rm eff}$, the gas locally falls into the
normal liquid phase and the effective distance from the phase transition point
increases (see dashed line (a) in Fig.~\ref{fig:phase-space}).

In order to check the theory at $\mu > 0$, we simulate the BH model at $\mu=2$.
The choice of this specific value for the chemical potential is driven by two
competing requirements:
on the one hand, we need $\mu$ be sufficiently large so that $T_c^\prime(\mu)$
be significantly different from zero, thus making the quadratic trapping
potential the dominant perturbation to the homogeneous system;
on the other hand, the different nature of the zero-temperature quantum phase
transition at the endpoint of the $\mu > 0$ transition line means that we must
keep $\mu$ sufficiently below $\mu=3$.
The value $\mu=2$ is in this sense a good compromise.

Thanks to the symmetry of the phase diagram of the HC model, the transition
temperature is known from previous works at $\mu = -2$ and its value reads
$T_c^{\mu=2} = 0.7410(1)$.
Contrary to the $\mu \leq 0$ case, moving out of the trap along a radius at 
$\mu > 0$, the system locally falls into the superfluid (low temperature)
phase (cf.~dashed line (b) in Fig.~\ref{fig:phase-space}).
At large distances, however, we also expect the system to cross again the
phase boundary on the opposite side of the phase diagram, i.e., when
$\mu_{\rm eff}(r) = -2$.

According to our previous considerations, we expect Eq. \ref{eq:xi_chi_ftss} to
hold at most as long as all the sites in the system belong to the same 
effective phase (in our case, the superfluid phase).
This requirement can be cast in the form
\begin{equation}
\mu_{\rm eff}\left( r_{\rm max} \right) > -2,
\end{equation}
where $r_{\rm max}$ is the distance of the farthest point from
the centre of the trap. 
In a 3D cube, $r_{\rm max} = \sqrt{3}L/2$, i.e., half of the length of the
diagonal of the box.
Using the definition for $\mu_{\rm eff}$, for a harmonic trap at $\mu = 2$, we
expect to observe finite-size and trap-size scaling behaviour at most up to
\begin{equation}
L_{\rm bd} = \frac{4l}{\sqrt{3}}.
\end{equation}

In Figure~\ref{fig:xi_chi_mu-2} we show the FTSS analyses on the data obtained
at the homogeneous critical temperature corresponding to $\mu=2$.
The QMC runs are approximately $1 \cdot 10^6$ MCS long.
The simulations substantially confirm the proposed scaling scenario.
The data of Fig.~\ref{fig:xi_chi_mu-2}-(a,b) collapse on a universal curve
characterized by the trap exponent $\theta_2$, even though sizeable corrections
to scaling are present (see discussion below).
At a sufficiently large value of $L/l^\theta$, the data fall out of the 
universal curve (see Fig.~\ref{fig:xi_chi_mu-2}-c). 
Quantifying the exact value of $L/l^\theta$ at which FTSS breaks down is a
difficult task, since we cannot sample the curve in more points. 
In our simulations, in fact, $L$ must be odd, so that the minimum step for the
data points in the figure is $2/l^{\theta}$ for any given $l$.
However, we can qualitatively say that the corresponding value of $L/l$ is
close to 2, which is in good agreement with the forecast value of
$4/\sqrt{3} \approx 2.3$.

We conclude this section by discussing in some detail the origin of the scaling
corrections.
We identify two main sources of corrections: the irrelevant operators already
present in the homogeneous system and the presence of the $O(V^2)$ term in
Eq.~\ref{eq:tau_mu}.
We can provide a rough quantitative estimate for the relative weights of
the $O(V)$ and $O(V^2)$ contributions by approximating the critical boundary
$T_c(\mu)$ with an ellipse with semi-axes fixed by the critical temperature at
$\mu = 0$ and the endpoint at $\mu = \pm 3$:
\begin{equation}
T_c(\mu) \approx T_c^{\mu=0} \sqrt{1 - \left( \frac{\mu}{3} \right)^2}.
\end{equation}
This ansatz reproduces the measured critical temperature at $\mu = 2$ within a
few percent.
We can then evaluate the coefficients $a$ of $V$ and $b$ of $V^2$ in
Eq.~\ref{eq:tau_mu} to find that they are of the same order of magnitude, with
$b/a \approx \frac{1}{2}$.
The scaling corrections due to the potential $V^2\sim (r/l)^{2p}$ are expected
to be\cite{PhysRevA.81.023606} of order
$O(l^{-2(1-\theta_2)}) \sim l^{-0.85\ldots}$ for $p=2$.
These must be compared with irrelevant perturbations of the homogeneous system,
in the presence of the harmonic potential alone. 
In the limit $L \gg l$, these are 
$O(l^{- \omega \theta_2}) \sim l^{-0.45\ldots}$,
whereas, for $L \gtrsim l$, they are $O(L^{-\omega}) \sim L^{-0.79\ldots}$.
We conclude that the irrelevant perturbations provide the largest
contribution to scaling corrections.

\section{Conclusions}\label{sec:conclusion}

We question the universality of the effective trap-size scaling theory first
proposed in Ref.~\onlinecite{PhysRevB.88.024517} by studying the critical 
behaviour of the trapped hard-core Bose-Hubbard model in 3D,
Eq.~\ref{eq:bh_trap}, at vanishing and positive chemical potential $\mu$.
The theory was so far only tested on the 2D Bose-Hubbard model, which belongs 
to the classical 2D XY universality class.

The standard TSS theory claims that the critical features of confined systems
close to the centre of the trap $V(r)\sim r^p$ is determined by a trap exponent
$\theta(p)$ that is shared among representatives of a common universality
class\cite{PhysRevLett.102.240601}.
The finite-$T$ quantum critical behaviour of the trapped BH model can be 
mapped\cite{PhysRevA.81.023606} onto that of the classical XY model trapped by 
a potential $U\sim r^q$.

The effective TSS theory builds upon these results by showing that the trapped
BH model at $\mu \neq 0$ corresponds to the XY model trapped by a potential 
with exponent $q=p$, while at $\mu = 0$ the correct mapping is $q = 2p$.
The different behaviour at vanishing chemical potential is due to the shape of
the superfluid lobe in the phase diagram of the model (see 
Fig.~\ref{fig:phase-space}), and in particular to the fact that 
\begin{equation}
 \left. \frac{\partial T_c(\mu)}{\partial \mu}\right|_{\mu=0} = 0,
\end{equation}
making the leading contributions of order $V^2\sim r^{2p}$.

To validate the theory, we simulated the homogeneous hard-core 3D BH model at
$\mu=0$ to determine the transition temperature. 
Our finite-size scaling analysis results in $T_c^{\mu=0} = 1.00801(4)$,
significantly improving the previous estimates.
We then simulated the trapped model at $\mu=0$ and $\mu=2$ (for which the
critical temperature was already known).

The data fully agree with the effective TSS theory.
At $\mu = 0$, our TSS analysis clearly favours the exponent $\theta(2p)$ over
$\theta(p)$, as predicted. 
At $\mu>0$ the exponent $\theta(p)$ rules the critical properties of the 
system, although the data suffer from strong corrections to scaling.
We detailed the sources of these corrections, and identified the dominant
contributions with those due to irrelevant perturbations already present in the
homogeneous system.

Notably, the effect of the trap on the system at positive $\mu$ is to locally
push the gas towards the superfluid phase.
Sufficiently far from the centre of the trap, the system may locally reach
$\mu_{\rm eff}(r) \leq -\mu$, thus crossing the phase boundary between the
superfluid and normal fluid phases.
When this happens, the two phases coexist inside the trap, leading to a break
down of TSS.

We remark that, although we focussed on the hard-core limit of the 
Bose-Hubbard model, our results extend to soft-boson systems through 
universality arguments.
In this sense, our work is relevant to experimental studies of cold bosonic
gases in optical lattices, in which these conditions may be concretely 
realised.
In these experiments, the momentum density distribution $n(\vec{k})$ is 
often measured.
This quantity is related to the two-point function by a Fourier transform,
\begin{equation}
 n(\vec{k}) = \sum_{\vec{x}, \vec{y}} e^{i\vec{k} \cdot (\vec{x} - \vec{y})}
      G_b(\vec{x}, \vec{y}),
\end{equation}
and is experimentally accessed by the analysis of absorption images after a 
time-of-flight.\cite{RevModPhys.80.885}
Unfortunately, $n(\vec{k})$ is not the ideal observable to probe the TSS 
critical behaviour.
In fact, only the correlations within approximately a distance 
$l^{-\theta}$ from the centre of the trap exhibit TSS 
scaling,\cite{PhysRevB.87.024513} whereas $n(\vec{k})$ integrates over all 
pairs of positions $(\vec{x}, \vec{y})$ in the lattice, thus suppressing the 
critical features by approximately a factor of the total volume of the system.

Instead, a more promising observable is the density-density correlation 
function relative to the centre of the trap,\cite{PhysRevB.87.024513}
\begin{align}
 G_n(\vec{0}, \vec{r}) 
       &= \avg{n_\vec{0}n_\vec{r}} - \avg{n_\vec{0}}\avg{n_\vec{r}} \nonumber\\
       &\approx l^{-2\theta(3 - 1/\nu)}\mathcal{G}_n(\vec{r}l^{-\theta}).
\end{align}
The latter is both accessible in 
experiments\cite{Nature.460.995, PhysRevA.81.031610, Science.329.547} via 
\emph{in situ} imaging of the atomic cloud and exhibits TSS criticality.
As already noticed at the end of Sec.~\ref{sec:TSS_mu0}, the fast depletion of 
the atomic cloud moving outwards from the centre of the trap requires that $l$ 
be sufficiently large in order for the signal to noise ratio of 
the correlations to be significant.
Furthermore, the trap size should be varied over a wide range of values in 
order to assess the critical scaling.

Despite the experimental challenges, any such measurement would constitute a 
significant leap from an approximate treatment of the confining potential to 
an exact probing of the influence of the trap on the critical behaviour of 
the system.

\subsection*{Acknowledgements}
We warmly thank E.~Vicari for his valuable advice and a critical reading of 
this manuscript. We also thank O.~Morsch for helpful discussion.
The QMC simulations and the data analysis were performed at the
\emph{Scientific Computing Center - INFN Pisa}.

\appendix
\section{Monte Carlo dynamics}
\label{sec:selfcorr}
\begin{figure}[b!]
 \centering
 \includegraphics[width=8.5cm]{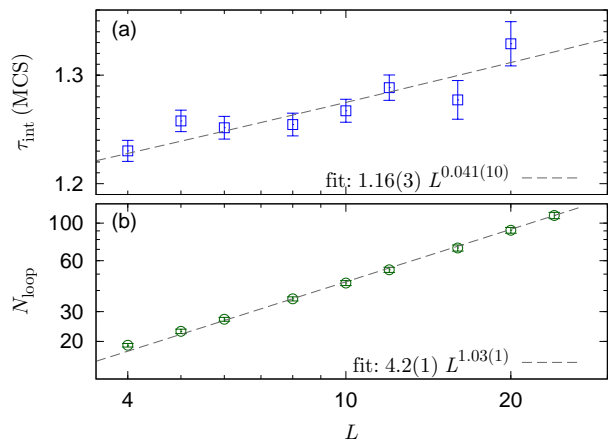}
 \caption{Bilograrithmic plots of the integrated self-correlation times          
          $\tau_{\rm int}(L)$ (a) and of the loop number $N_{\rm loop}$ (b).}
 \label{fig:ctimes}
\end{figure}

The SSE with directed loops is a MC algorithm which acts on an extended
$(d+1)$-dimensional configuration space.
The Hamiltonian of the system is written in terms of diagonal and off-diagonal
bond operators.
These operators, together with the identity operator, are inserted along the
extra dimension of the configuration space.
\cite{PhysRevE.66.046701, PhysRevA.85.053637}
A MCS is divided in three phases:
(i) diagonal update (DU), in which diagonal and identity operators may be
swapped;
(ii) off-diagonal update (ODU), during which $N_{\rm loop}$ loops are built and
diagonal and off-diagonal operators are exchanged with each other in the
configuration;
(iii) free-spin flipping (FSF), during which the sites of the $d$-dimensional
lattice upon which no operator acts are flipped randomly.

$N_{\rm loop}$ is determined during equilibration and kept constant during the
run.
At fixed physical parameters, it may however vary slightly as the seed of the
random number generator is changed.
In our simulations, we coarsely round $N_{\rm loop}$ so that in all the runs at
given physical parameters it takes the same value.
From Fig.~\ref{fig:ctimes}-b we get an almost linear dependence of
$N_{\rm loop}$ as $L$ increases.

We estimate the scaling properties of the MC dynamics for the homogeneous
system by looking at the integrated self-correlation time $\tau_{\rm int}$ of
the critical observable $\Upsilon$ at the critical temperature \ref{eq:tc_mu0}.
In general
\begin{equation}\label{eq:tau_int}
 \tau_{\rm int} \propto L^d L^z,
\end{equation}
where $z$ is the dynamical exponent of the MC.
From Fig.~\ref{fig:ctimes}-a we observe that, in units of MCS, $\tau_{\rm int}$
is almost constant as $L$ increases.

However, it must be kept in mind that the MCS is not an elementary update, but
is made of one DU, followed by $N_{\rm loop}$ loops and finally one FSF.
At the critical temperature, the computational effort for all the elementary
updates (DU, loop, FSF) scales as the volume of the system.
Moreover, the time needed for the DU and the FSF is negligible compared with
the time of the ODU.
According to Eq.~\ref{eq:tau_int} and to the evidence of Fig.~\ref{fig:ctimes},
we conclude that the dynamical exponent is $z \approx 1$.

\vfill
\bibliography{notes.bib}

\end{document}